# Spin-half paramagnetism in graphene induced by point defects

R. R. Nair[1], M. Sepioni[1], I-Ling Tsai[1], O. Lehtinen[2], J. Keinonen[2], A. V. Krasheninnikov[2,3], T. Thomson[1], A. K. Geim[1], I. V. Grigorieva[1*]

[1]Manchester Centre for Mesoscience & Nanotechnology, Manchester M13 9PL, UK
[2]Department of Physics, P.O. Box 43, FI-00014 University of Helsinki, Finland
[3]Department of Applied Physics, P.O. Box 11100, Aalto University, FI-00076, Finland

*Using magnetization measurements, we show that point defects in graphene – fluorine adatoms and irradiation defects (vacancies) – carry magnetic moments with spin 1/2. Both types of defects lead to notable paramagnetism but no magnetic ordering could be detected down to liquid helium temperatures. The induced paramagnetism dominates graphene's low-temperature magnetic properties despite the fact that maximum response we could achieve was limited to one moment per approximately 1000 carbon atoms. This limitation is explained by clustering of adatoms and, for the case of vacancies, by losing graphene's structural stability.*

The possibility to induce magnetic response in graphene by introduction of defects has been generating much interest, as in principle this could allow the addition of controlled magnetism to the already impressive list of special properties of this material, offering possibilities for novel devices where charge and spin manipulation could be combined. To date there have been many theoretical studies (for review, see [1-3]) predicting that point defects in graphene, such as vacancies and adatoms, should carry a magnetic moment $\mu \sim \mu_B$ and these can in principle couple (anti)ferromagnetically [1-13]. On the other hand, experimental evidence of defect-induced magnetic moments in graphene remains both scarce and controversial [14-17], unlike for the case of graphite. Indeed, the emerging consensus that magnetism in carbon-based systems can exist is based on a large body of work on magnetic measurements of highly-oriented pyrolytic graphite (HOPG) and carbon films, with many reports of weak room-temperature ferromagnetic signals observed in both pristine HOPG and after its ion irradiation (see, for example, [18-21]). However, the whole subject remains controversial, especially concerning (i) the role of possible contamination and (ii) the mechanism responsible for the strong interaction required to lead to ferromagnetic ordering at room temperature (*T*). Some observations of ferromagnetism are probably artifacts (one frequent artifact is identified and described in the Supplementary Information, where we show that most commonly used HOPG crystals contain micron-size magnetic particles). Adatom magnetism in graphite is also contentious and, for example, different studies of fluorinated graphite reported inconsistent results [22,23].

Graphene, as the basic building block of all graphitic materials, is simpler than its 3D counterparts and can be used to address questions which are harder to address in the 3D materials with complex electronic spectra. When sufficient quantities of graphene became available in the form of laminates (large collections of electronically non-interacting graphene crystallites), it has become possible to use SQUID magnetometry to show that pristine graphene is strongly diamagnetic and shows only tiny background paramagnetism that requires *T* <50 K to be noticeable [16]. As concerns the role of defects in graphene, the most relevant observations to date are probably the report of spin-1 paramagnetism in ion-implanted graphitic nanoflakes [17] and the Kondo effect and giant negative magnetoresistance in defected graphene devices [24,25], although the former disagrees with the spin ½ universally expected for point defects [1-13] and the latter are transport experiments allowing different interpretations [13,25]. Let us also mention the observation of a peak in the tunneling density of states near vacancies, even though that work was carried out for graphite [26].

In this report, we have employed the above graphene laminates as a well-characterized and clean, reference system to study the effect of controlled introduction of point defects on graphene's magnetic properties. Two types of defects have been studied: (i) fluorine (F) adatoms in concentrations *x* gradually increasing to stoichiometric fluorographene $CF_{x=1.0}$ [27] and (ii) vacancies introduced by ion irradiation. Both types of defects lead to notable paramagnetism, significantly exceeding the ferromagnetic fraction often reported for 3D graphitic compounds. In the case of adatoms, the maximum achievable concentration of paramagnetic centers in our work was $\sim 10^{-3}$ per carbon atom, much smaller than fluorine concentrations. We attribute this relative inefficiency to fluorine clustering, so that neighboring adatoms residing on different graphene's sublattices do not contribute to the overall magnetic moment. For vacancies, their density was limited by the requirement to retain the structural integrity of graphene.

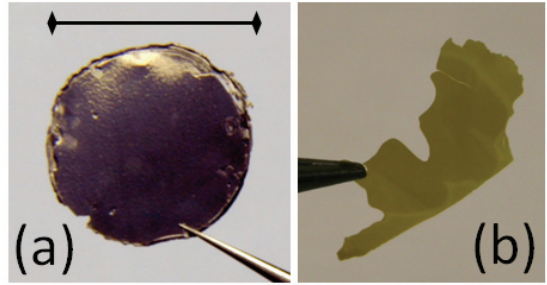

FIGURE 1. Optical images of (a) initial and (b) fully fluorinated graphene laminates. The scale bar is 2 cm.

Starting samples were prepared by ultrasonic cleavage of high-purity HOPG in an organic solvent, N-methylpyrrolidone, following the procedure described in ref. [16] and Supplementary Information. The resulting laminates (Fig. 1a) consist of 10 to 50 nm graphene crystallites, predominantly mono- and bi-layers, aligned parallel to each other and electronically decoupled. Magnetization behavior of the pristine laminates was described in detail in our previous study [16]. Briefly, it shows a purely diamagnetic response over a wide $T$ range. Weak background paramagnetism becomes noticeable only below 50K and corresponds to a low density (~40ppm) of magnetic moments, with approximately one spin per graphene crystallite. The origin of this paramagnetism remains unclear but is unlikely to be due to remnant magnetic impurities or graphene's edge states and point defects [16]. To achieve a gradually increasing concentration of fluorine atoms attached to graphene sheets, we followed the procedures of ref. [27], namely, a whole sample such as that shown in Fig.1a was placed in a Teflon container and, for a number of hours, exposed to atomic F formed by decomposition of xenon difluoride at 200°C. During fluorination graphene was in contact only with $XeF_2$ and Teflon, and test measurements (mock run) showed that no contamination was introduced during our preparation procedures. A progressive increase in the F/C ratio was monitored by using Raman spectroscopy (Supplementary Information), as well as the color change: from metallic dark grey, through brown to light yellow - see Fig. 1b. For quantitative analysis we used X-ray photoemission spectroscopy, which allowed us to measure the value of $x$ in $CF_x$ (e.g. $CF_{x=0.1}$ corresponds to an F/C ratio of 0.1). Details of the composition analysis are given in Supplementary Information.

The exposure of the graphene laminates to atomic fluorine resulted in strong paramagnetism, such that the low-$T$ saturation magnetization increased by more than an order of magnitude with respect to the background signal in initial samples. Figure 2 shows the evolution of the magnetization with increasing $x$ from 0.1 to 1 (fluorographene). At all fluorine concentrations, the behavior can be accurately described by the Brillouin function

$$M = NgJ\mu_B \left[ \frac{2J+1}{2J} \text{ctnh}\left(\frac{(2J+1)z}{2J}\right) - \frac{1}{2J} \text{ctnh}\left(\frac{z}{2J}\right) \right]$$

where $z = gJ\mu_B H / k_B T$, $g$ is the $g$-factor, $J$ the angular momentum number, $N$ the number of spins and $k_B$ the Boltzmann constant. The Brillouin function provides good fits only for $J=S=1/2$ (free electron spin). Other $J$ unequivocally disagree with the functional form of the measured $M(H)$ as they give qualitatively different, sharper changes with quicker saturation (for details of defining $J$, see Supplementary Information). This behavior is corroborated by fits of M(T) to the Curie law curves $\chi = M/H = NJ(J+1)g^2\mu_B^2/(3k_BT)$, which were calculated for $J=1/2$ and $N$ inferred from the $M(H)$ curves – see Fig. 2b. Given the many reports of defect-induced ferromagnetism in carbon systems, we paid particular attention to any signs of magnetic ordering in our samples. No signatures were found with accuracy better than $10^{-5}$ emu/g – see inset in Fig. 2b.

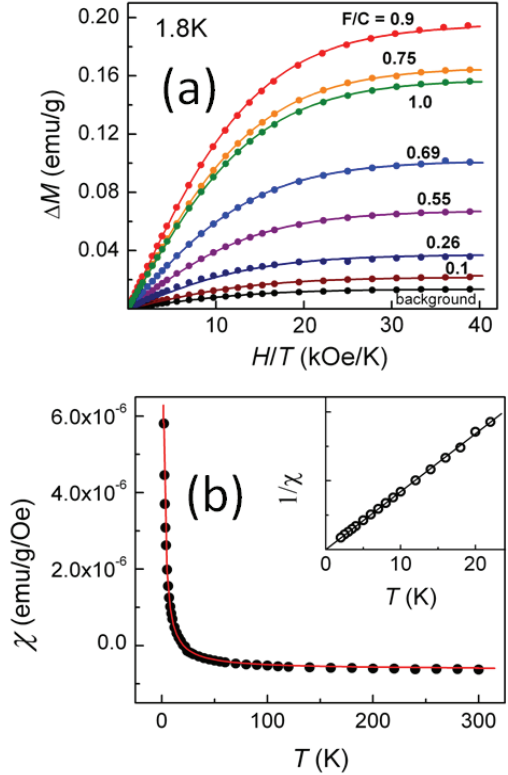

FIGURE 2. Paramagnetism due to fluorine adatoms. (a) Magnetic moment $\Delta M$ (after subtracting a diamagnetic background) as a function of parallel field $H$ for different F/C ratios. Symbols are the measurements and solid curves are fits to the Brillouin function with $S=1/2$ and assuming $g=2$ (there are no indications in literature that $g$-factor in graphene may be enhanced). (b) Example of $T$ dependence of susceptibility $\chi=M/H$ in parallel $H=3$kOe for $x=0.9$; symbols are the measurements and the solid curve is the Curie law calculated self-consistently using the $M/H$ dependence found in (a). Inset: Inverse susceptibility vs $T$ demonstrating the linear, purely paramagnetic behavior with no sign of magnetic ordering.

The evolution of the number of spins $N$, extracted from the measured $M(H,T)$, with an increasing degree of fluorination is shown in Fig. 3. $N$ increases monotonically with $x$ up to $x\approx 0.9$, and then shows some decrease for the fully fluorinated samples. It is instructive to replot the same data in terms of the number of Bohr magnetons, $\mu_B$, per attached F atom (inset in Fig. 3). It is clear that the initial increase (up to $x\approx 0.5$) in the number of paramagnetic centers is proportional to $x$, as expected, but for higher $x$ the behavior reveals a more complicated relation between the number of adatoms and $N$. Furthermore, for all $x$ the measured number of paramagnetic centers is three orders of magnitude less than the measured number of F adatoms in the samples, i.e., only 1 out of ~1000 adatoms appears to contribute to the paramagnetism. This may seem surprising because the general expectation is that each adatom should contribute $\sim\mu_B$ to the total $M$ [1-13]. However, we recall that fluorine atoms on graphene have a strong tendency towards clustering due to low migration barriers [28-30]. The tendency towards clustering is further enhanced in the presence of corrugations (ripples) [28-31]. The value of $\mu_B$/F_atom of $\sim 10^{-3}$ implies clusters of ~8 nm in size, which agrees with the measured typical sizes of ripples in graphene [31]. In the case of clustering, the magnetic moment from the interior of a cluster is expected to be zero because of the bipartite nature of the graphene lattice [1,3,32]. Therefore, a magnetic contribution can come only from clusters' edges and would be determined by a particular configuration of adatoms near the edges, i.e., only those adatoms on the A sublattice that have no counterparts on the neighboring sites of the B sublattice will contribute to *M*. Accordingly, each cluster is expected to contribute only a few $\mu_B$ to the total magnetization, which can explain our observations.

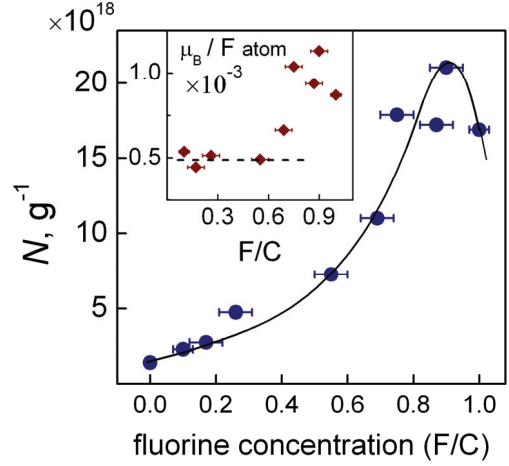

FIGURE 3. Adatom paramagnetism. Main panel: Number of spins *N* extracted from the Brillouin fits in Fig. 2 as a function of F/C ratio. The solid curve is a guide to the eye. Inset: the same *N* normalized to the concentration of adatoms in each sample ($\mu_B$/F atom is obtained by dividing the number of moments *N*, assuming that each carries $1\mu_B$, by the number of F atoms per g of fluorinated graphene).

As *x* increases, it is reasonable to expect that at some moment all strongly curved areas of the graphene sheet become occupied with fully-formed F clusters so that there appears an increasing number of isolated (that is, magnetic) adatoms or small clusters. Their contribution can explain the observed non-linear increase in $\mu_B$ per adatom for *x* >0.5. As $CF_x$ approaches the stoichiometric compound (*x*=1), we observe a notable (~20%) decrease in *M* as compared to *x* ≈0.9 (confirmed in several independent measurements) but the material remains strongly paramagnetic. The latter is somewhat surprising because stoichiometric fluorographene should probably [12] be nonmagnetic. However, even for *x*=0.999, there are still a large number of defects present in the fluorographene lattice (~ one missing F atom per 1000) and we speculate that the resulting midgap states associated with such defects and found in fluorographene [27] can have a magnetic moment, similar to the case of graphane [12].

To explore the generality of the magnetic behavior described above, we investigated magnetic response associated with another type of point defects, namely, vacancies produced by irradiation of graphene with high-energy protons and carbon ($C^{4+}$) ions. Unlike adatoms, vacancies are not mobile at room *T* and cannot form clusters but otherwise their contribution to magnetism is expected to be similar to that of adatoms (both types of defects lead to the formation of localized states at energies corresponding to the vanishing density of states [1-12] and are often referred to as $p_z$-vacancies [1]). Also, note that interstitial carbon atoms created along with the vacancies are not expected to carry a magnetic moment [33].

For our irradiation experiments, we started with the same graphene laminates as used for fluorination. The energy, fluence and other irradiation parameters for each sample were chosen on the basis of computer simulations (SRIM software package) to achieve a desired defect density and to ensure uniformity of defect distribution. The ion energies – 350-400 keV for protons and 20 MeV for $C^{4+}$ - were chosen to ensure that the stopping ranges for the $H^+$ and $C^{4+}$ ions exceeded the sample thicknesses, thus leaving behind only irradiation-induced defects but no implanted ions. During the irradiation, special care was taken to prevent sample contamination and to avoid any significant vacancy-interstitial recombination by ensuring that the sample temperature remained below 50°C. The total number of vacancies in each sample was estimated by using SRIM simulations. Further details are given in Supplementary Information.

The results of magnetization measurements for vacancies are summarized in Fig. 4. For all concentrations, *M*(*H*) curves, such as those shown in the inset, are found to be described accurately by magnetism of non-interacting spins with $S=1/2$ (solid curves are fits to the Brillouin function). This result disagrees with ref. [17] but is in agreement with theory and directly proves the case of much discussed magnetism due to vacancies in graphene [1-13]. Also, it is clear from Fig. 4 that irradiation defects give rise to the same paramagnetism, independent of the kind of ions used ($H^+$ or $C^{4+}$), and there is no sign of ferromagnetic ordering (also, see Supplementary Information).

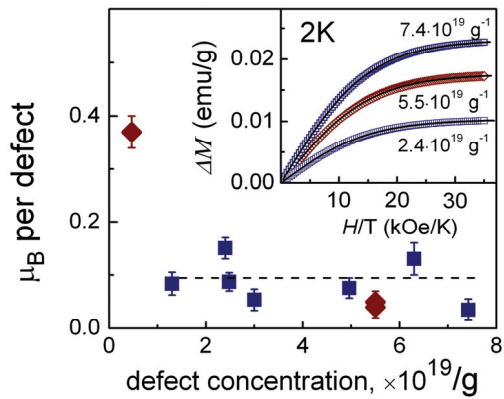

FIGURE 4. Vacancy paramagnetism. Magnetic properties of graphene laminates irradiated with protons (blue squares) and $C^{4+}$ ions (red diamonds). Main panel: Magnetic moment $M$ normalized to the concentration of vacancies. Inset: $M$ as a function of $H$. The labels give the defect density; solid curves are Brillouin function fits for $J=1/2$ (Supplementary Informartion).

At first glance, the effect of irradiation is rather similar to that of fluorination. However, the maximum total increase in $M$ that we could achieve in the irradiation experiments was ~10 times smaller than the maximum $M$ achieved with adatoms (cf. Fig. 2). This is simply because we could only introduce a limited amount of vacancies and, for defect densities above $10^{20}$ g$^{-1}$, our samples became so fragile that they disintegrated. Despite the smaller achieved magnetization, the effect of *individual* vacancies is much larger than in the case of F adatoms. Fig. 4 shows the magnetic moment per vacancy, calculated from the $M(H)$ curves, versus the estimated density of vacancies in each of the studied samples. The found values of $\mu$ lie between 0.1 and 0.4 $\mu_B$ per vacancy, which should be considered in reasonable agreement with the theoretical value of $\mu \sim \mu_B$ expected for isolated vacancies [1-13]. Indeed, there are considerable uncertainties in estimating the number of vacancies in our samples. First, SRIM simulations work well for graphite but cannot take into account the exact crystallography of graphene laminates and the varying separations between crystallites. More importantly, there is a possibility of reconstruction of individual vacancies into, for example, double vacancies, chemically terminated dangling bonds, Stone-Wales defects, etc. which are expected to be nonmagnetic [11]. In addition, the possibility of vacancy-interstitial recombination during or after irradiation cannot be ruled out completely (Supplementary Information).

In conclusion, we have demonstrated that vacancies and adatoms in graphene have spin 1/2. This provides the most unambiguous and direct support so far for many theories discussing graphene's magnetic properties. As for intensively-debated ferromagnetism in graphitic compounds, no magnetic ordering was detected, with both fluorinated and irradiated samples exhibiting purely paramagnetic behavior even for the largest defect densities at lowest $T$ of 2K. This agrees with straightforward expectations. Indeed, the concentration of magnetic moments we could achieve was only 0.1% of the maximum hypothetically possible magnetism of one moment per carbon atom. In the case of adatoms, this relative inefficiency of introducing paramagnetic centers is explained by clustering, so that neighboring adatoms residing on different graphene's sublattices do not contribute to the overall magnetic moment. For vacancies, their density is limited by the requirement to retain the structural integrity of graphene. Although the achieved concentrations of paramagnetic centers are notably higher than the ferromagnetic fractions typically reported for graphitic materials, an average spacing between magnetic moments is still rather large ~8 nm, apparently too large for magnetic ordering to take place even at liquid-helium *T*.

## Supplementary Information

**#1 Quantitative analysis of fluorination of graphene laminates**

Fluorination of graphene laminates was studied using different characterisation techniques such as Raman spectroscopy, X-ray photoelectron spectroscopy (XPS), and energy dispersive X-ray microanalysis (EDX). Raman spectroscopy provided a quick qualitative analysis of different levels of fluorination. This showed that the evolution of Raman spectra of graphene paper with fluorination (not shown) was very similar to the previously reported spectra for mechanically exfoliated single layer graphene [S1]. For quantitative determination of the fluorine-to-carbon ratio (F/C) after different fluorination times we employed XPS. Furthermore, the XPS results for several samples were corroborated by EDX.

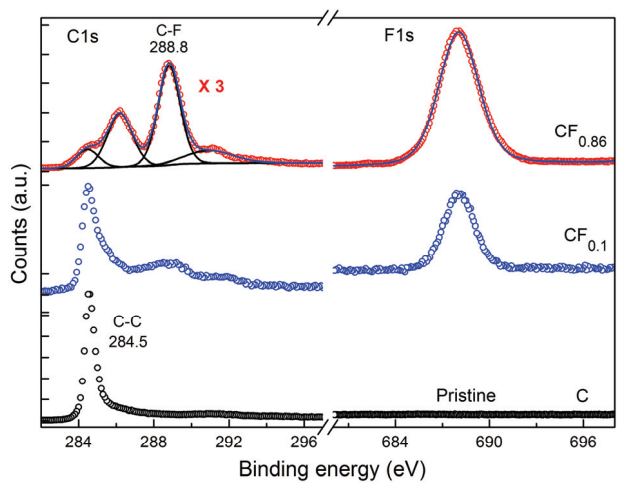

**Figure S1**. Typical examples of XPS spectra for pristine (bottom curve) and fluorinated (middle and top curves) graphene laminates at different degrees of fluorination.

XPS spectra were acquired on a Kratos Axis Ultra X-ray photoelectron spectrometer equipped with a monochromatic aluminium X-ray source and a delay line detector. Typical wide scan spectra were recorded from a 220 μm diameter area for binding energies between 0 and 1100eV with a 0.8eV step, at 160eV pass energy. High resolution spectra of C1s and F1s peaks (such as those in Fig. S1) were recorded with a 0.1eV step, at 40 eV pass energy.

Figure S1 shows typical XPS spectra for energies corresponding to C1s peaks of carbon (left part) and F1s peaks of fluorine (right part). The three C1s peaks indicate the presence of different types of carbon bonds while the strong F1s peak appearing after fluorination indicates the presence of chemisorbed fluorine. An intense peak at ≈ 284eV in the spectrum of pristine graphene paper is due to the C-C $sp2$ bonds in graphene [S1]. After exposure of graphene laminates to fluorine the intensity of this peak decreases while new peaks appear, corresponding to different types of C-F bonding. The intensities of these new peaks vary for different degrees of fluorination. The most pronounced peak at ≈289eV, in conjunction with the strong fluorine peak at ≈688eV, yield strong C-F covalent bonding [S2]. As Fig. S1 shows, the ≈688eV peak is absent for pristine graphene laminates but shows strongly for all fluorinated samples, which again confirms the presence of covalently bonded fluorine. The other two peaks at ≈286.3eV and 291eV indicate the presence of C-CF and $CF_2$ bonds, respectively [S1, S3]. $CF_2$ and $CF_3$ bonds are expected to form primarily at defects or edges of graphene flakes. Note that the absence of $CF_3$ peaks (≈ 293eV) and very small intensity of the $CF_2$ peaks indicates that fluorination in our samples occurs mainly at graphene surfaces, rather than at the defect sites or edges. As all graphene crystallites in the laminates are very small (typically 30-50 nm), bi- and trilayer graphene flakes (that are present in our laminates along with ~50% monolayers) are easily intercalated by F atoms.

The stoichiometric composition of different fluorinated samples was obtained by analysing the intensity of carbon and fluorine peaks using Casca XPS software [S4]. Spectra were taken from several ≈100 μm spots on each sample and the results averaged to obtain F/C for a particular fluorinated sample. This procedure indicated good spatial homogeneity of fluorination on a scale »100 μm, as the calculated F/C ratios varied by no more than 5%. In addition, we carried out EDX analysis of several fluorinated samples, to confirm the stoichiometry, and found that the results were in full agreement with the XPS analysis above. These showed that for the longest fluorination time (80h) we were able to achieve F/C ≈1, i.e. stoichiometric fluorographene. EDX analysis was also used to confirm that our fluorinated samples were free from contamination with metal impurities.

**#2 Irradiation of graphene laminates with high-energy ions**

For $C^{4+}$ irradiations we used 5MV van de Graaf tandem accelerator where samples, mounted on aluminium foil, were clamped to a copper rod cooled with liquid nitrogen to -50°C. The sample temperature, which was monitored using an infrared temperature sensor, remained below 50°C during irradiation and the vacuum was maintained in the $10^{-6}$-$10^{-7}$ mbar range. To ensure irradiation homogeneity, the focused beam spot was rasterized over the sample area using two perpendicular electromagnets. The fluence was determined from the total charge accumulated in the target chamber.

Proton irradiations were carried out in a 500kV ion implanter, at room temperature and energies between 350 and 400 keV and current densities <0.2 µA/cm$^2$. Two perpendicular electric fields were used to sweep the beam over the sample area and the radiation fluence was measured using four Faraday cups with a known window area. No considerable heating of the target is expected with the used energies and currents. The concentration of backscattered protons which stopped in the sample is estimated to be less than 1 ppm.

The fluences used to achieve the defect densities in Fig. 4 of the main text varied from $5\cdot10^{13}$ to $1\cdot10^{16}$ cm$^{-2}$. We note that it was not possible to derive the defect density for each sample from the fluence alone, because different samples had different surface areas and different defect distributions. The defect densities shown in Fig. 4 (main text) were calculated for each sample on the basis of the corresponding fluence, surface area and the number of vacancies created by each type of ions as obtained from SRIM simulations.

To check for the possibility of vacancy-interstitial recombination during or after irradiation [S5], one of the proton-irradiated samples was annealed at 300°C for 8h, which resulted in ~20% reduction in magnetic moment. This shows that the majority of vacancies in our samples do not recombine even at temperatures much higher than room temperature, in agreement with the higher mobility of interstitials in few-layer graphene compared to graphite [S6].

**#3 Determination of the spin value from magnetisation data**

To characterize the magnetic species contributing to the magnetization $M$ of fluorinated and irradiated graphene laminates, we plotted $M$ as a function of the reduced field $H/T$. For all samples (i.e. all degrees of fluorination and all vacancy concentrations) $\Delta M(H/T)$ curves measured at different temperatures collapse on a single curve, indicating that graphene with both types of defects behaves as a paramagnet with a single type of non-interacting spins. As an example, Fig. S2(a) shows such analysis for the fully fluorinated graphene $CF_{x=1}$. The observed behavior is accurately described by the Brillouin function, such that the initial slope of $\Delta M(H/T)$ is determined by the angular momentum quantum number $J$ and the $g$ factor, and the saturation level is determined by the number of magnetic moments (spins) $N$:

$$M = NgJ\mu_B\left[\frac{2J+1}{2J}\text{ctnh}\left(\frac{(2J+1)x}{2J}\right) - \frac{1}{2J}\text{ctnh}\left(\frac{x}{2J}\right)\right]$$

where $x = gJ\mu_B H/k_B T$ and $k_B$ the Boltzmann constant. Assuming $g=2$ (there are no indications in literature that g-factor in graphene may be enhanced), the Brillouin function provides excellent fits to the data for $J = S = 1/2$ (red curve Fig. S2(a)). For comparison, we also show fitting curves for $J = 1, 3/2$ and 2, all of which provide very poor fits, making it clear that only $J = S = 1/2$ fits the data.

Figure S2(b) shows similar analysis for two different vacancy concentrations in graphene laminates irradiated with 350 keV protons. Again, only $J=S=1/2$ fits the data, with all other values of $J$ giving very poor fits. This provides the most unequivocal proof that both types of point defects studied - fluorine adatoms and vacancies - represent non-interacting paramagnetic centers with spin $S=1/2$.

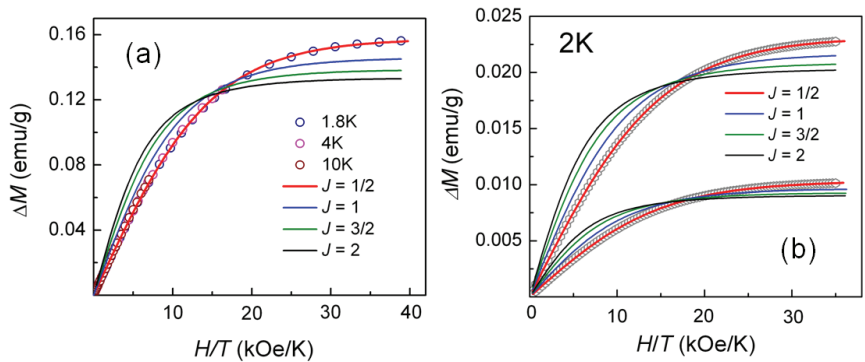

**Figure S2**. Determination of the angular momentum quantum number $J$ from $\Delta M$ vs $H/T$ curves. (a) Magnetisation of the fully fluorinated graphene crystallites, $x$=1 (linear diamagnetic background subtracted). Symbols are data for three different temperatures and solid curves are fits to the Brillouin function for $J$ =1/2, 1, 3/2 and 2. (b) Magnetisation due to vacancies: As the increase in $M$ after irradiation was comparable to the paramagnetic signal in pristine graphene, $\Delta M$ here is the magnetisation over and above the paramagnetic contribution measured before irradiation (linear diamagnetic background subtracted as well). Grey symbols show data for two different vacancy concentrations (lower curve $2.4 \cdot 10^{19}$ g$^{-1}$, upper curve $7.4 \cdot 10^{19}$ g$^{-1}$) and solid curves show Brillouin function fits with $g$=2 and $J$ =1/2, 1/, 3/2 and 2.

#### #4 Commentary on ferromagnetism reported for HOPG

Weak ferromagnetic signals (~$10^{-3}$ emu/g) were found in pristine highly-oriented pyrolytic graphite (HOPG) (e.g. [S7,S8]) which, according to the authors, could *not* be explained by ~1-2ppm of Fe detected using particle-induced X-ray emission (PIXE) or X-ray fluorescence spectroscopy (XRFS). Accordingly, the ferromagnetism was attributed to intrinsic defects, such as, e.g., grain boundaries [S8]. The ferromagnetic response was shown to increase dramatically after high-energy ion irradiation of HOPG [S9-S13], nanodiamonds [S14], carbon nanofoams [S15] and carbon films [S16]. Several scenarios have been suggested to explain the observed ferromagnetism.

Trying to clarify the situation, we have carried out extensive studies of magnetic behaviour of HOPG crystals obtained from different manufacturers (ZYA-, ZYB-, and ZYH-grade from NT-MDT and SPI-2 and SPI-3 from SPI Supplies). These crystals are commonly used for studies of magnetism in graphite; e.g., ZYA-grade crystals were used in refs. S7, S9-S13 and ZYH-grade in ref. S8. We have also observed weak ferromagnetism, similar in value to the one reported previously for pristine (non-irradiated) HOPG. Below, we show that the ferromagnetism in ZYA-, ZYB-, and ZYH-grade crystals is due to micron-sized magnetic inclusions (containing mostly Fe), which can easily be visualized by scanning electron microscopy (SEM) in the backscattering mode. Without the intentional use of this technique, the inclusions are easy to overlook. No such inclusions were found in SPI crystals and, accordingly, in our experiments these crystals were purely diamagnetic at all temperatures (no ferromagnetic signals at a level of $10^{-5}$ emu/g).

Ten HOPG crystals of different grades (ZYA, ZYB, ZYH and SPI) were studied using SQUID magnetometry (Quantum Design MPMS XL7), XRFS, SEM and chemical microanalysis by means of energy-dispersive X-ray spectroscopy (EDX). For all ZYA, ZYB and ZYH crystals, magnetic moment vs field curves, $M(H)$, showed characteristic ferromagnetic hysteresis in fields below 2000 Oe, which was temperature independent between 2K and room $T$, implying a Curie temperature significantly above 300K. The saturation magnetisation $M_S$ varied from sample to sample by more than 10 times, from $1.2 \cdot 10^{-4}$ emu/g to $3 \cdot 10^{-3}$ emu/g – see Fig. S3. This is despite the fact that XRFS did not detect magnetic impurities in any of our HOPG crystals (with a detection limit better than a few ppm). This result is similar to the findings of other groups [e.g. S9, S10, S11, S16]. Figure S3 also shows an $M(H)$ curve for one of the SPI crystals, where no ferromagnetism could be detected.

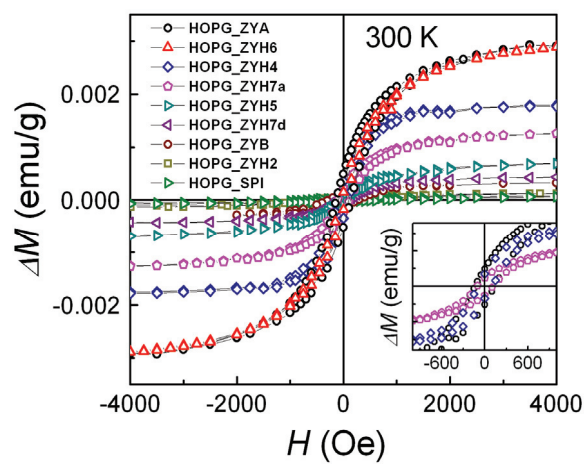

**Figure S3**. Ferromagnetic response in different HOPG crystals. Magnetic moment $\Delta M$ vs applied field $H$ after subtraction of the linear diamagnetic background. The inset shows a low-field zoom of three curves from the main panel where the remnant $\Delta M$ and coercive force are seen clearly.

The seemingly random values of ferromagnetic signal in nominally identical crystals could be an indication that the observed ferromagnetism is related to structural features of HOPG, such as grain boundaries, as suggested in ref. [S8]. However, we did not find any correlation between the size of the crystallites making up HOPG crystals and/or their misalignment and the observed $M_S$. For example, the largest $M_S$ as well as the largest coercive force, $M_C$, were found for one of the ZYA crystals, which have the smallest mosaic spread (0.4-0.7°), and for a ZYH crystal with the largest mosaic spread (3-5°). Furthermore, crystallite sizes were rather similar for all ZYA, ZYB and ZYH crystals (see Fig. S4) while $M_S$ varied by almost a factor of 3 (see Fig. S3).

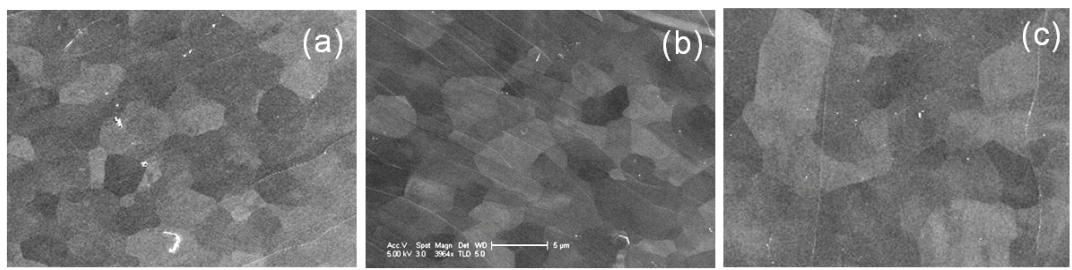

**Figure S4**. Typical, same-scale, SEM images of crystallites in different HOPG samples: (a) ZYH; (b) ZYA; (c) SPI. Typical crystallite sizes in ZYH, ZYB and ZYA are 2 to 5 μm; in SPI crystallites vary from 0.5 to 15 μm. The scale bar corresponds to 5μm.

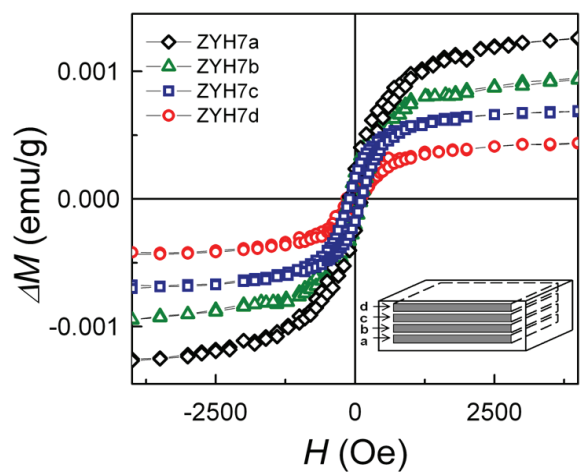

**Figure S5**. Ferromagnetic hysteresis in four samples cut from the same ZYH HOPG crystal. The inset shows schematically positions of the samples in the original crystal.

To investigate whether the observed ferromagnetism is homogeneous within the same commercially available 1cm×1cm×0.2cm HOPG crystal, we measured magnetisation of four samples cut out from the same ZYH crystal as shown in the inset of Fig. S5. To exclude possible contamination of the samples due to exposure to ambient conditions, both exposed surfaces were cleaved and the edges cut off just before the measurements. Surprisingly, we found significant variations of the ferromagnetic signal between these four nominally identical samples – see Fig. S5. This indicates that the observed ferromagnetism is not related to structural or other intrinsic characteristics of HOPG crystals, as these are the same for a given crystal. Therefore, it seems reasonable to associate the magnetic response with external factors, such as, for example, the presence of small inclusions of another material.

To check this hypothesis, we examined samples of different HOPG grades using backscattering SEM. Due to its sensitivity to the atomic number [S17], backscattered electrons can provide a strong contrast allowing to detect particles made of heavy elements inside a light matrix (graphite in our case). This experiment revealed that all ZYA, ZYB, and ZYH crystals contained sparsely distributed micron-sized particles of a large atomic number, with typical in-plane separations of 100 to 200 µm – see Fig. S6. Comparison of SEM images in backscattering and secondary electron modes (BS and SE, respectively) revealed that in most cases the particles were buried under the surface of the sample and, therefore, were not visible in the most commonly used secondary electron mode. This is illustrated in Fig. S7 which shows the same area of a ZYB sample in the SE and BS modes.

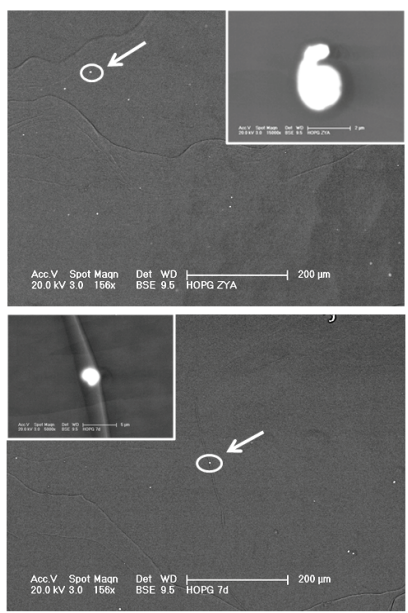 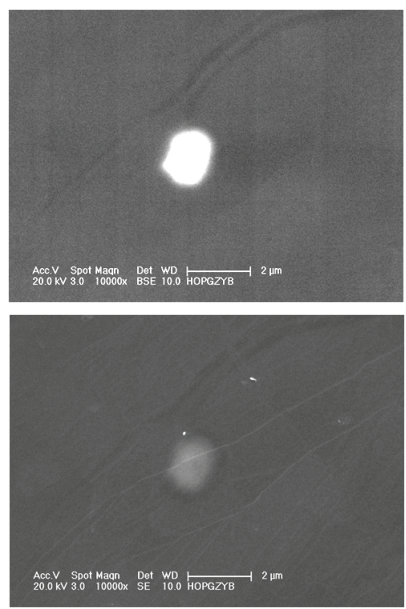

**Figure S6.** SEM images of ZYA (top) and ZYH (bottom) samples in back-scattering mode. Small white particles are clearly visible in both images, with typical separations between the particles of 100 µm for ZYA and 240 µm for ZYH. Insets show zoomed-up images of the particles indicated by arrows; both particles are ≈2µm in diameter.

**Figure S7**. SEM images of the same particle found in a ZYB sample taken in backscattering (top) and secondary electron (bottom) modes. Surface features are clearly visible in the SE image while BS is mostly sensitive to chemical composition. The contrast around the particle in the SE mode is presumably due to a raised surface in this place.

The difference between the two images is due to different energies and penetration depths for secondary and backscattered electrons: the energy of BS electrons is close to the primary energy, i.e. ~20 keV in our case, and they probe up to 1µm thick layer at the surface [S17] while secondary electrons have characteristic energies of the order of ~50 eV and come from a thin surface layer of a nm thickness [S18]. Importantly, no such inclusions could be detected in SPI samples that, as discussed, did not show any ferromagnetic response.

The difference between ZY and SPI grades is presumably due to different manufacturing procedures used by different suppliers. Our attempts through NT-MDT to find out the exact procedures used for production of ZY grades of HOPG were unsuccessful.

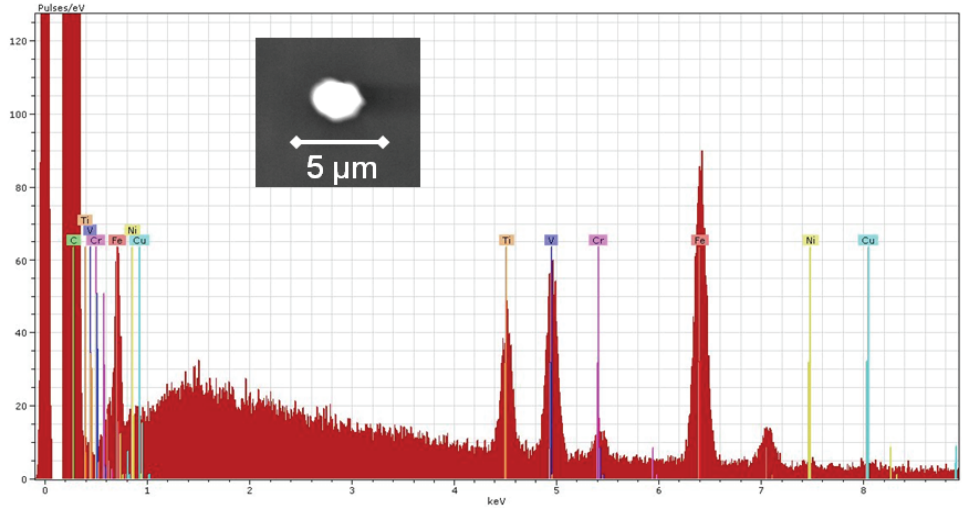

**Figure S8.** EDX spectrum of one of the particles found in a ZYA sample. Inset shows SEM image of the particle.

To analyse the chemical composition of the detected particles we employed *in situ* energy-dispersive X-ray spectroscopy (EDX) that allows local chemical analysis within a few μm$^3$ volume. Figure S8 shows a typical EDX spectrum collected from a small volume (so-called interaction volume) around a 2.5 μm diameter particle in a ZYA sample. This particular spectrum corresponds to the presence of 8.6 wt% (2.1 at%) Fe, 2.3 wt% (0.65%) Ti, 1.8 wt% V (0.47 at%) and <0.5 wt% Ni, Cr and Co, as well as some oxygen, which appear on top of 86 wt% (96.5 at%) of carbon. The latter contribution is attributed to the surrounding graphite within the interaction volume. To determine the actual composition of the inclusion, we needed to take into account that the above elemental analysis applies to the whole interaction volume, where the primary electrons penetrate into the sample. Given that 96% of the interaction volume is made up by carbon, the electron range *R* and, accordingly, the interaction volume can be estimated to a good approximation using the Kanaya-Okayama formula [S19]:

$$R = \frac{0.0276 \cdot A \cdot E^{1.67}}{Z^{0.89} \cdot \rho} \approx 4.5 \mu m,$$

where *A*=12 g/mole is the atomic weight of carbon, *E*=20 keV the beam energy, *Z*=6 the atomic number and ρ=2.25 g/cm$^3$ the density of graphite. Using the calculated value of *R*, the weight percentages for different elements from the spectrum and their known densities, it is straightforward to show that the volume occupied by the detected amount of Fe and Ti is in excellent agreement with the dimensions of the particle in Fig. S7, i.e., the particle is made up predominantly of these two elements. The presence of oxygen indicates that Fe and Ti are likely to be in an oxidised state, i.e. the particle is either magnetite or possibly titanomagnetite, both of which are ferrimagnetic, with saturation magnetisation $M_S \approx$ 75-90emu/g [S20].

We estimate that a 2.5μm diameter particle of magnetite contributes ≈2.5·10$^{-9}$ emu to the overall magnetisation. Therefore, the observed ferromagnetic signal (1.5·10$^{-5}$ emu) for this particular ZYA sample (3×3×0.26 mm) implies that the sample contains ~6,000 magnetite particles which, if uniformly distributed, should be spaced by ~100 μm in the *ab* plane. This is in agreement with our SEM observations. This allows us to conclude that the visualized magnetic particles can indeed account for the whole ferromagnetic signal for this sample.

BS and EDX analysis of the other HOPG samples showing ferromagnetism produced similar results, with some samples containing predominantly Fe and others both Fe and Ti, as in the example above. A clear correlation has been found between the value of $M_S$ for a particular sample and the average separation of the magnetic particles detected by BS/EDX – see Fig. S9. No magnetic particles could be found in SPI samples and, accordingly, they did not show any ferromagnetic signal.

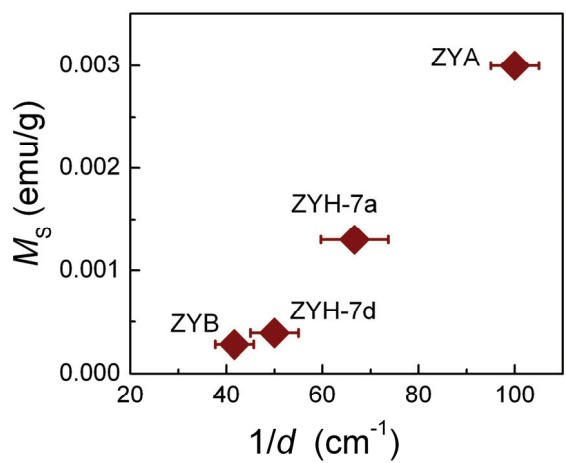

**Figure S9.** Saturation magnetisation $M_s$ as a function of the inverse of the average separation between the detected particles $d$ as determined from the backscattering images (see text).

On the basis of the above analysis, we conclude that ferromagnetism in our ZYA, ZYB and ZYH HOPG samples is not intrinsic but due to contamination with micron-sized particles of probably either magnetite or titanomagnetite. As these particles were usually detected at submicron distances below the sample surface (see above), they should have been introduced during high-temperature crystal growth. We note that ZYA, ZYB or ZYH grades of HOPG are most commonly used for studies of magnetism in graphite (e.g., ZYA-grade crystals were specified in refs. [S7, S9-S13]) and, therefore, the contamination could be the reason for the often reported ferromagnetism.

Finally, we would like to comment why magnetic particles such as those observed in the BE mode could have been overlooked by commonly used elemental analysis techniques, such as XRFS and PIXE [S7-S16]. Assuming that all particles found in our samples are magnetite and of approximately the same size, 2-3 μm, we estimate that the total number of Fe and Ti atoms in our samples ranges from 1 to 6 ppm. In the case of XRFS, 5 ppm of Fe is close to its typical detection limit and these concentrations might remain unnoticed. In the case of PIXE, its resolution is better than 1ppm. PIXE was used in e.g. refs. [S9, S11] for ZYA graphite, where no contamination was reported but the saturation magnetisation was ≈ $(1-2) \cdot 10^{-3}$ emu/g, similar to our measurements. The absence of detectable concentrations of magnetic impurities has been used as an argument that the ferromagnetic signals could not be due to contamination. Also, it was usually assumed that any remnant magnetic impurities were distributed homogeneously, rather than as macroscopic particles, and therefore would give rise to paramagnetism rather than a ferromagnetic signal, which was used as an extra argument against magnetic impurities.

It is clear that the latter assumption is incorrect, at least for the case of ZY grade graphite. Furthermore, let us note that the difference of several times for the limit put by PIXE and the amount measured by SQUID magnetometry is not massive. In our opinion, this difference can be explained by the fact that PIXE tends to underestimate the concentration of magnetic impurities if they are concentrated into relatively large particles. Indeed, PIXE probes only a thin surface layer (~1μm for 200 keV protons) which is thinner than the diameter of the observed magnetic inclusions. One can estimate that for round-shape inclusions with diameters of ~3μm, there should be a decrease by a factor of 3 in the PIXE signal with respect to the real concentration. Even more importantly, inclusions near the surface of HOPG provide weak mechanical points and are likely to be removed during cleavage when a fresh surface is prepared. Therefore, we believe that a micron-thick layer near the HOPG surface is unlikely to be representative of the whole sample. In contrast, magnetisation measurements probe average over the bulk of the samples, which can explain the observed several times discrepancy.

**#5 Remnant ferromagnetism in graphene laminates**
Preparation of graphene laminates involves splitting of graphite into individual graphene planes. Therefore, if standard HOPG ZYH or ZYA crystals are used, one can expect that magnetic particles present in the starting crystals may in principle pass into graphene laminates, despite the fact that centrifugation should mostly remove heavy particles from the suspension. Indeed, careful measurements of sufficiently large samples of graphene laminates made from ZYH-grade HOPG detected very weak ferromagnetic signals – see Fig. S10. For a typical 1.5-2mg graphene laminate sample the ferromagnetic component of magnetisation, $M_S$ (300K), was found to be $<5 \cdot 10^{-7}$ emu. Therefore, measuring the corresponding hysteresis loops required the maximum possible sensitivity of our SQUID magnetometer ($<1 \cdot 10^{-7}$ emu), which we achieved by using the RSO option and taking great care in terms of sample mounting and background uniformity. Apart from the smaller $M_S$, all other characteristics of the ferromagnetic signal in the laminates (temperature independence, remnant magnetisation and coercive force) were similar to those for the starting HOPG samples, indicating the same origin of ferromagnetism – see Fig. S9.

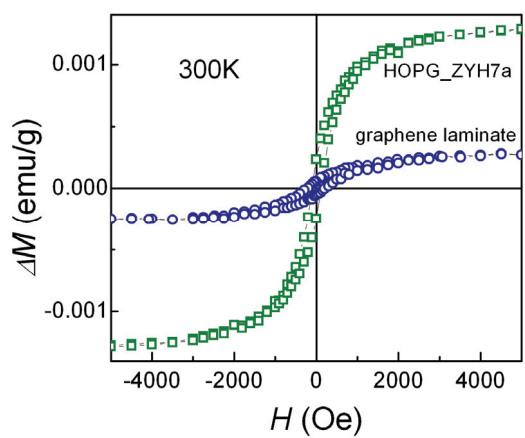

**Figure S10**. Comparison of the ferromagnetic magnetisation in HOPG ZYH and a typical sample of graphene laminate

Some samples used in our fluorination and irradiation experiments were additionally purified by immersion for several hours in a mixture of $HNO_3$ and HF acids and subsequent annealing at 300°C. The above acids are efficient in dissolving metal impurities, either in oxidised or pure metal form. After such treatment, no ferromagnetism was observed in any of these samples. Similar treatment of HOPG samples led to a reduction in $M_S$ but not complete disappearance of the ferromagnetic signal. This result is easy to understand, as acids can penetrate between the decoupled graphene layers in laminate samples relatively easily, and this is not the case for bulk HOPG.

We conclude that the very weak but still detectable ferromagnetic response that we sometimes observed in graphene samples prepared from commonly available HOPG crystals also results from the initial contamination of HOPG, which should be taken into account when studying magnetic properties of graphene laminates and its derivatives.